\documentclass[aps,prl,twocolumn,showpacs,floatfix,letterpaper,longbibliography,superscriptaddress]{revtex4-2}

\usepackage{graphicx}
\usepackage[english]{babel}
\usepackage{bm}
\usepackage{amsmath,amsfonts}
\usepackage{amsbsy}
\usepackage[colorlinks=true,linkcolor=blue,urlcolor=blue,citecolor=blue,breaklinks=true]{hyperref}
\usepackage[utf8]{inputenc}

\usepackage{graphics}
\usepackage{subcaption}
\usepackage{xcolor}
\usepackage{bbm}
\usepackage[normalem,normalbf]{ulem}

\usepackage{changes}
\usepackage{ulem}
\usepackage{soul}

\usepackage{amssymb}
\usepackage{array}
\usepackage{amsmath}
\usepackage{graphicx}\usepackage{graphics}
\usepackage{dcolumn}
\usepackage{bm}\usepackage{varioref}
\newcommand{\RNum}[1]{\uppercase\expandafter{\romannumeral #1\relax}}

\begin{document}


\title{
Electronic pumping of heat without charge transfer
}

\author{A. V. Andreev}
\affiliation{Department of Physics, University of Washington, Seattle, WA 98195,  USA}
\affiliation{Skolkovo  Institute of  Science  and  Technology,  Moscow,  143026,  Russia}
\affiliation{L. D. Landau Institute for Theoretical Physics, Moscow, 119334 Russia}

\date{\today}

\begin{abstract}
A mechanism of electron-mediated pumping of heat in the absence of net charge transfer is proposed.  It may be realized in charge-neutral electron systems, such as graphene, coupled to an external electric potential. The flow of heat in this pumping cycle is not accompanied by a buildup of voltage along the system, which offers advantages over traditional thermoelectric cooling setups.  Efficiency of heat pumping and magnitude of heat flux are studied in the hydrodynamic regime for weak disorder. In a pristine system, even for an infinitesimal pumping potential the heat flux remains finite. In particular, for a  potential in the form of a traveling wave moving with velocity $c$ the pumping is perfect; the entire heat content of the electron liquid is advected with velocity $c$. For a general pumping cycle the  heat flux is determined by the cycle geometry and disorder strength.
\end{abstract}


\maketitle

\addtolength{\abovedisplayskip}{-1mm}

In typical heat pumping cycles the heat is transferred between the hot and cold reservoirs convectively by some physical substance - the working body, or coolant.
Convection enables rapid transfer of heat between spatially separated reservoirs. In contrast, non-convective spreading of heat relative to physical substances typically proceeds via a much slower diffusion process.

However, diffusive character of spreading of heat through matter is not universal; under some conditions propagation of heat through physical substances may become ballistic. The well known examples are second sound waves~\cite{khalatnikov1989introduction} in superfluid $^4$He~\footnote{Recently,  ballistic spreading of heat was also predicted~\cite{Matveev_second_sound} to occur in quantum one-dimensional liquids at low temperature.},  and pure crystals in the regime of phonon hydrodynamics~\cite{gurevich1986transport}.  In contrast to the usual adiabatic sound, in which entropy and matter move together,  the temperature oscillations in the second sound wave are practically decoupled from the density oscillations. They correspond to temperature/entropy waves that move ballistically through matter.  Crystallization  waves in $^4$He~\cite{andreev1978equilibrium} represent another  example of decoupling between the flows of order and matter. In particular, a fall of a $^4$He crystallite in superfluid $^4$He proceeds partly via melting at the top and crystallization at the bottom of the crystallite~\cite{tsymbalenko2019effect}. As a result the crystalline order propagates at a faster speed than the matter itself.

Ballistic propagation of heat can also be realized in electronic systems~\cite{phan2013ballistic} in  the regime of electron hydrodynamics~\cite{Gurzhi:1968,de1995hydrodynamic,andreev2011hydrodynamic}.
In systems with equal densities of electrons and holes (e.g. in graphene at charge neutrality) the hydrodynamic flow of the electron liquid is decoupled from charge flow and corresponds to the flow of heat, see Ref.~\cite{Lucas_2018} for a recent review.  Hydrodynamic regime of charge-neutral electron liquid has been realized recently in monolayer and bilayer graphene~\cite{Crossno1058,tan2019realization}, and may  be realized in clean semimetals.

Ballistic spreading of heat through matter presents an enticing possibility to design novel heat pumps, in which pumping of heat is accomplished in the absence of matter flow. Such a pumping cycle requires an ability to induce and guide the flow heat through the system by coupling it to external perturbations. In this respect electronic systems seem particularly promising because flow of electrons can be readily controlled by time-dependent gate voltages, thus obviating the need for mechanically moving parts.

Pumping of charge in microelectronic devices has been extensively studied for more than four decades. This work was largely stimulated by Thouless' prediction of quantization of pumped charge in adiabatic quantum pumps~\cite{Thouless_pump}.  Quantization of pumped charge in Coulomb blockade devices~\cite{Kouwenhoven_1991,Pothier_1992} is used in metrological applications, e.g. the modern realization of capacitance standard~\cite{Martinis_charge_pump,Keller_capacitance_2011}.  Pumping of charge in open mesoscopic systems~\cite{Brouwer_1998,Aleiner_1998,Spivak_1999,Switkes1999}, and its accuracy~\cite{FCS_pump_2000,FCS_Makhlin_2001,FCS_Coulomb_blockade_2001,FCS_Nazarov_2003}  were also extensively studied.

The linear coupling between the heat and charge current, i.e. the Peltier effect~~\cite{abrikosov2017fundamentals}, enables using electron pumps for heat  pumping.  Thermoelectric effects in microelectronic devices in Coulomb blockade regime~\cite{beenakker1992theory,Staring_1993,Matveev_thermopower} and in the regime of electron hydrodynamics~\cite{andreev2011hydrodynamic,Lucas_2018,levchenko2010transport,degottardi2015electrical,li2020hydrodynamic}  have been studied.
The goal of this article is to show that decoupling of heat and charge flows in electron liquids at charge neutrality may be utilized to  create heat pumps, which transfer heat in the absence of net flow of charge.

Let us consider a pumping setup, in which heat transfer is mediated by an electron liquid in a charge-neutral system, such as  monolayer or bilayer  graphene,  subjected to an external electric pumping potential $U({\bm r}, t)$. The latter may be generated by applying time-dependent voltage to a series of gates, or for example,  using method developed in Ref.~\cite{Talyanskii_2005} to achieve quantized charge pumping in carbon nanotubes -- by placing the system on a piezoelectric substrate driven by a surface acoustic wave (SAW).

The essential features of the pumping mechanism  can be understood most easily for clean systems in the regime of electron hydrodynamics.  Consider a unidirectional geometry with a periodic pumping potential of the form of a traveling wave, $U(x - c t)$. At zero temperature difference between the reservoirs, and for slow pumping velocities $c$,  the electron liquid will remain in local thermal equilibrium corresponding to the instantaneous realization of the pumping potential $U(x - c t)$.  Therefore the densities of electrons, $n(x,t)$ and entropy, $s(x,t)$, will be given by the equilibrium values corresponding to the local value of $U$.   Since the latter  moves with velocity $c$,  the electron liquid will also move with the hydrodynamic velocity $u =c$. The local densities of charge and entropy of the electron liquid will propagate with the same velocity. Therefore, the entire heat content of the electron liquid will be entrained by this flow, producing a net heat flux  density $T\overline{ s} c $   where $\overline{\cdots}$ denotes spatial average. In contrast,  the net  charge pumping current  will vanish because of the vanishing average electron density at charge neutrality, $\overline{n} =0$.

In the presence of disorder  and temperature gradient the entrainment of the electron liquid by the pumping potential will no longer be prefect; the pressure gradient proportional to the temperature gradient and the disorder-induced friction force will cause the electron liquid to lag behind the pumping potential. Evaluating the heat flux and pumping cycle efficiency in this case requires a quantitative theory.

Below, a theory of heat pumping at charge neutrality is developed in the regime of electron hydrodynamics. The hydrodynamic description applies provided  the rate of  momentum-conserving  electron-electron collisions exceeds the momentum relaxation rate and the pumping frequency $\omega$.   Let us assume that the spatial scale of the pumping potential exceeds the  correlation radius $\xi$ of the disorder potential. In this case pumping may be described by averaging the flow of the electron liquid over length scales of order $\xi$ ~\cite{andreev2011hydrodynamic,Lucas_2018,li2020hydrodynamic}.  For slow pumping the corresponding  macroscopic hydrodynamic equations may be written in the form
\begin{subequations}\label{eq:hydro_average_draft}
\begin{eqnarray}
\label{eq:hydro_n}
  \partial_t n + \bm{\nabla}\cdot \bm{j}&=& 0, \\
  \label{eq:hydro_p}
  \partial_t \bm{p} + k \bm{u} + \bm{\nabla} P + n \bm{\nabla} (U + e\phi)&=& \bm{\nabla} \cdot \hat{\sigma}' ,\\
  \partial_t s + \bm{\nabla}\cdot \bm{j}_s &=& \dot{s} .\label{eq:hydro_s}
\end{eqnarray}
\end{subequations}
Here $P$ is the pressure, $\bm{u}$ is the hydrodynamic velocity, and $\hat{\sigma}'$ denotes the viscous stress tensor,  and $k$ denotes the disorder-induced ``friction'' coefficient. The densities of particles, entropy, and momentum are denoted by
 $n$, $s$, and $\bm{p}$,  respectively. The electric potential $\phi$ is related to the electron charge density $e n$ by the Poisson equation. The  current densities of particles, $\bm{j}$, and entropy, $\bm{j}_s$ may be expressed as
\begin{equation}\label{eq:currents}
\left(
                      \begin{array}{l}
                        \mathbf{j} \\
                        \mathbf{j}_s \\
                      \end{array}
                    \right)
                    =
  \left(
    \begin{array}{c}
      n \\
      s \\
    \end{array}
  \right) \bm{u} -
   \left(
                                 \begin{array}{cc}
                                   \sigma/e^2 & \gamma/T \\
                                   \gamma/T & \kappa/T  \\
                                 \end{array}
                               \right)   \left(
    \begin{array}{c}
      - e\boldsymbol{\mathcal{E}} \\
       \boldsymbol{\nabla} T \\
    \end{array}
  \right)
  ,
\end{equation}
where the first term in the right hand side (r.h.s.) represents the equilibrium components of the currents, while the second represents  the dissipative components. The latter are linear in the  temperature gradient $\bm{\nabla}T$ and the electromotive force (EMF) $e\bm{\mathcal{E}} = - \bm{\nabla}(\mu + U + e\phi)$ (with $\mu$ being the chemical potential).  The elements of the Onsager matrix of the intrinsic kinetic coefficients of the electron liquid are the electrical and thermal conductivities $\sigma$, and $\kappa$, and  thermoelectric coefficient $\gamma$. Finally,    $\dot{s}$ is entropy production rate per unit area, caused by dissipative processes in the electron liquid and loss of heat to the lattice.

\emph{Adiabatic pumping.} Let us focus on the regime of slow pumping, where the heat flux is linear in the rate of change $\partial_t U$ of the pumping potential, and further assume that the temperature difference $\Delta T$ between the hot and cold reservoirs is small. In this case we may work within linear order accuracy in $\partial_t U$ and $\Delta T$.  Note that momentum density $\bm{p}$ is linear in these variables. Since its time derivative is further proportional to the pumping rate,  $\partial_t \bm{p}$ is quadratic in $\partial_t U$ and $\Delta T$. Similarly, the entropy production rate $\dot{s}$ is also quadratic in these variables. Therefore, in our approximation we may neglect the force density $\partial_t \bm{p}$ and entropy production rate $\dot{s}$ in Eq.~\eqref{eq:hydro_average_draft}, thereby reducing the entropy evolution equation, Eq.~\eqref{eq:hydro_s}, to a continuity relation for the entropy current.

Let us consider a  unidirectional geometry, in which a two-dimensional system of length $L$  (in the $x$-direction) and width $w$ (in the $y$-direction) is subjected  to a  periodic in space and time pumping potential of the form $U(x,t) = U(x + \lambda, t) = U(x, t + \tau)$. Being interested in the bulk effects we will evaluate the pumping heat flux per unit width of the system for  $L\gg \lambda$. In this case without loss of generality we will set $L/\lambda$ to be an integer, and impose at the reservoirs periodic boundary conditions on the system variables,   $e\mathcal{E}$, $\partial_x T$, and $u$.

Using the  thermodynamic identity $d P = n d\mu + s d T$ and expressing the relevant component of the viscous stress tensor as $\sigma_{xx}'= (\eta + \zeta) \partial_x u$, where $\eta$ and $\zeta$ are the shear and bulk viscosities, we obtain from  Eq.~\eqref{eq:hydro_p} the following force balance relation
\begin{align}\label{eq:force_balance}
	n e\mathcal{E} - s\partial_x T - \left[k - \partial_x(\eta + \zeta)\partial_x \right] u  =& 0 .
\end{align}
The remaining two relations  between $u$, $e\mathcal{E}$, and $\partial_x T$ are obtained by
integrating continuity equations \eqref{eq:hydro_n} and \eqref{eq:hydro_s}, over $x$, and using Eq.~\eqref{eq:currents};
\begin{subequations}\label{eq:hydro_int}
	\begin{eqnarray}
		\label{eq:hydro_int_n}
		n u + \sigma_0 e \mathcal{E}  - \frac{\gamma \partial_x T}{T} &=& j(t) - \int_0^x d \tilde{x} \partial_t n(\tilde{x},t) , \\
		s u -\frac{\kappa \partial_x T}{T} + \frac{\gamma e \mathcal{E}}{T}  &=& j_s (t) - \int_0^x d \tilde{x} \partial_t s(\tilde{x},t) .\label{eq:hydro_int_s}
	\end{eqnarray}
\end{subequations}
Here the integration constants $j(t)$ and $j_s(t)$ represent, respectively, the current densities of particles and entropy evaluated at the reservoirs, $x=0$ and $x=L$ (due to the periodic boundary conditions the two are equal).

Equations \eqref{eq:force_balance} and \eqref{eq:hydro_int} determine the hydrodynamic velocity $u$,  temperature gradient $\partial_x T$, and EMF $e\mathcal{E}$, which arise in the presence of pumping. They do not assume that pumping potential is weak, only that the pumping, described by the integrals in the r.h.s. of Eq.~\eqref{eq:hydro_int} is adiabatically slow. In the general case of strong pumping potential not only the densities of particles and entropy, but also the kinetic coefficients depend on the pumping potential, and are given by their locally equilibrium values~\cite{andreev2011hydrodynamic}. These equations
must be supplemented by the boundary conditions for the temperature and electrochemical potential at the reservoirs. At charge neutrality they are given by $\overline{e \mathcal{E}} =0$ (vanishing voltage bias), and $\overline{\partial_x T} = \Delta T/L$.

For simplicity, let us assume that both pumping potential $U$ and disorder are small in comparison to $T$. We will work to lowest order accuracy in $U/T\ll 1$. To this end we will approximate all quantities to leading order in $U/T$. Since the local electron density $n$ is linear in the pumping potential, $n/s \propto U/T \ll 1$. The thermoelectric coefficient $\gamma$, being odd in $n$ is also linear in $U$.
In contrast,  deviations of the entropy density from its value  at charge neutrality, $s_0$, are quadratic  in $U/T$ and may be neglected. Similarly, the intrinsic electrical and thermal conductivities $\sigma$ and $\kappa$ may be replaced by their values at charge neutrality. Then it follows from Eq.~\eqref{eq:hydro_int_s} that to within linear order accuracy in $U/T$ the entropy flux $j_s$ is spatially uniform, $
j_s =j_s(t)= s_0 u -\kappa \partial_x T/T + \gamma e \mathcal{E}/T$.  Excluding $\partial_x T$ from Eqs.~\eqref{eq:force_balance} and \eqref{eq:hydro_int_s} we get
\begin{align}\label{eq:u_E}
  \left[\frac{T s_0^2}{\kappa} + k - \partial_x (\eta + \zeta)\partial_x \right] u  + \left[ \frac{\gamma  s_0 }{\kappa } - n \right] e\mathcal{E} & = \frac{Ts_0 j_s(t)}{\kappa} .
\end{align}
At charge neutrality the spatially uniform component of the EMF vanishes,  $\overline{e \mathcal{E}} =0$, whereas the inhomogeneous part of $e \mathcal{E}$ is $\propto U/T$. Therefore, it follows from the above equation that to within linear order accuracy in $U/T$ the hydrodynamic velocity is spatially uniform. Furthermore, the force balance equation \eqref{eq:force_balance} shows that to linear order in $U/T$ the temperature gradient is also spatially uniform, $\partial_x T = \Delta T/L$, and is related to the hydrodynamic velocity $u(t)$ by,
\begin{equation}\label{eq:zero_force_1}
	k u (t)  = \overline{n e \mathcal{E}} - s_0 \frac{\Delta T}{L}.
\end{equation}
The second  term in the r.h.s. describes the force density caused by thermally induced pressure gradient. This force is balanced by the disorder-induced friction force in the left hand side (l.h.s.) and the force exerted by the pumping potential (first term in the r.h.s.). This force may be evaluated by finding the local EMF from Eq.~\eqref{eq:hydro_int_n}. Isolating the the inhomogeneous part of Eq.~\eqref{eq:hydro_int_n} we obtain within our accuracy,
\begin{align}
	\label{eq:EMF_pumping}
	e\mathcal{E}(x,t)= -\frac{e^2}{\sigma_0}\left[ n(x,t) u (t)    +\int^x_0  d \tilde{x}\partial_t n(\tilde{x},t) \right].
\end{align}

Substituting Eq.~\eqref{eq:EMF_pumping} into  \eqref{eq:zero_force_1} and expressing the friction coefficient in terms of the variance $\langle (\delta n)^2\rangle$ of disorder-induced density modulations, $k =\frac{\sigma_0}{ 2e^2}\, \langle (\delta n)^2\rangle $~\cite{Crossno1058,li2020hydrodynamic}
we obtain the hydrodynamic velocity in the form
\begin{align}
	\label{eq:u_result}
	u (t)& = -  \frac{   \frac{ \sigma_0}{e^2} \frac{s_0 \Delta T}{L}  +   \overline{   n (x,t) \int^x_0  d \tilde{x}\partial_t n (\tilde{x},t)    }    }{\frac{\langle (\delta n)^2\rangle}{2} +  \overline{n^2(x,t)}}.
\end{align}
The entropy current may be expressed in terms  $u(t)$ using Eq.~\eqref{eq:u_E}. To leading (zeroth) order in $U/T$ one obtains
\begin{align}\label{eq:j_s}
	j_s (t) = s_0 u (t),
\end{align}
which corresponds to the flow of entropy density $s_0$ of the electron liquid with velocity $u(t)$.

Equations \eqref{eq:u_result} and \eqref{eq:j_s} describe the flow velocity $u(t)$ of the electron liquid and entropy flux in the presence of pumping and temperature difference between the reservoirs. The flow is caused by the thermally-induced pressure gradient $s_0 \Delta T/L$ (first term in the numerator in Eq.~\eqref{eq:u_result}) and the force exerted on the liquid by the pumping potential (second term).

At  $\Delta T = 0$ the flow velocity $u(t)$ is determined only by the pumping cycle parameters and disorder strength. This pumping contribution is given by
\begin{align}
	\label{eq:u_pure_pumping}
	u_\mathrm{P} (t)& = -  \frac{     \overline{   n (x,t) \int^x_0  d \tilde{x}\partial_t n (\tilde{x},t)    }    }{\frac{\langle (\delta n)^2\rangle}{2} +  \overline{n^2(x,t)}} .
\end{align}
In particular, for a potential in the form of a traveling wave generated by SAW, $U(x,t) = U_0(x-ct)$ we get $u^{\mathrm{SAW}}_\mathrm{P}  = c \left(  1  + \langle (\delta n)^2\rangle/2 \overline{n^2}   \right)^{-1} $.
At vanishing disorder this expression reproduces the expected result of perfect pumping, $u^{\mathrm{SAW}}_\mathrm{P} = c$.

At weak disorder, $ \langle (\delta n)^2\rangle  \ll   \overline{n^2(x,t)} $, the pumping  velocity in Eq.~\eqref{eq:u_pure_pumping} becomes independent of the amplitude of the pumping potential for a general pumping cycle. In other words it depends only on the shape of that pumping cycle $U(x,t)$.  Let us denote this value by  $u^{(0)}_\mathrm{P} (t)$.  Using  the Fourier series representation for the electron density, $n(x,t)= \sum_n c_n (t) e^{i \frac{2\pi n }{\lambda}  x}$,  the pumping velocity $u^{(0)}_\mathrm{P} (t)$    can be expressed as
\begin{align}
	\label{eq:u_pure_pumping_Fourier}
	u^{(0)}_\mathrm{P}(t) &=  \frac{ - i \sum_n \frac{\lambda}{2 \pi n }c^*_n(t) \partial_t c_n(t)    }{\sum_n |c_n(t)|^2},
\end{align}
and may be interpreted as adiabatic connection~\cite{shapere1989geometric} in the space of periodic functions with zero mean.

Averaging the $u(t)$ in Eq.~\eqref{eq:u_result} over time we can write the average pumping heat flux in Eq.~\eqref{eq:j_s} in the form
\begin{align}
	\label{eq:heat_flux_average}
	T\, j_s  & = T s_0 \overline{\overline{u}}_\mathrm{P}  - \kappa_{\mathrm{eff}}\frac{\Delta T}{L}, 
\end{align}
where $\quad \overline{\overline{u}}_\mathrm{P} = \frac{1}{\tau} \int_0^\tau dt  	u_\mathrm{P} (t)$ is the time-averaged velocity, and
 $\kappa_{\mathrm{eff}}$   represents the effective thermal conductivity of the system,
\begin{align}
	\label{eq:kappa_eff}
	\kappa_{\mathrm{eff}} = \frac{T \sigma_0 s_0^2 }{e^2} \,    \frac{1}{\tau} \int_{0}^{\tau}  \frac{d t }{\frac{\langle (\delta n)^2 \rangle}{2} + \overline{n^2(x,t)} }.
\end{align}
It is inversely proportional to the variance  of deviations of electron density from charge neutrality, which are caused by both disorder and the pumping potential. The factor $1/2$ arises because disorder-induced density variations, $\langle (\delta n)^2 \rangle$, are statistically isotropic, whereas the density variations induced by the pumping potential depend only on the $x$-coordinate.   At vanishing pumping potential  Eq.~\eqref{eq:kappa_eff} reproduces the result of Ref.~\cite{Crossno1058}.

Equations \eqref{eq:u_pure_pumping}, \eqref{eq:heat_flux_average},  and \eqref{eq:kappa_eff} describe  the average heat flux across the system.      At $\Delta T = 0 $ the pumping heat current is given by the first term in the r.h.s  of Eq.~\eqref{eq:heat_flux_average}. The pumping velocity of the electron liquid depends only on the pumping potential and disorder strength. At weak disorder, $\langle (\delta n)^2 \rangle \ll \overline{n^2}$, the pumping velocity in Eq.~\eqref{eq:u_pure_pumping} becomes independent of the amplitude of pumping potential, and is determined only by the pumping cycle geometry, see Eq.~\eqref{eq:u_pure_pumping_Fourier}. The flow of heat due to $\Delta T \neq 0$ is described by the second term in the r.h.s. of \eqref{eq:heat_flux_average} and is proportional to the effective thermal conductivity of the system in Eq.~\eqref{eq:kappa_eff}. The latter also depends on the pumping potential.  Note that, similar to peristaltic pumps, the pumping potential is distributed throughout the system. Therefore, the pumped heat flux (first term in the r.h.s. of Eq.~\eqref{eq:heat_flux_average})  does not decrease with the system length  $L$. In contrast, the flow of heat caused by the temperature difference $\Delta T$ (second term in the r.h.s. of Eq.~\eqref{eq:heat_flux_average}) is inversely proportional to  $L$. Therefore, for a fixed $\Delta T$, cooling may be achieved even at low pumping strengths for sufficiently long systems.

Let us now discuss the heat pump efficiency. To keep the expressions simpler we consider the traveling wave potential of the form $U_0(x- ct)$.  Efficiency of heat pumps is characterized by the coefficient of performance (COP), defined as the ratio of useful power to the power $\dot{W}$ consumed by the pump. For cooling/heating cycle the useful power (per unit width of the system) is given by $ Tj_s = T s_0 u$, where $T$ is the temperature of cold/hot reservoir.  The power $\dot{W}$ consumed by the pump may be obtained from a mechanical consideration.  Using Eq.~\eqref{eq:EMF_pumping} it is easy to see that for a traveling wave potential,  $U_0(x-c t)$,  the   force density exerted by the pumping potential on the electron liquid in Eq.~\eqref{eq:zero_force_1} has the form of a friction force, $\overline{n e \mathcal{E}} = \frac{e^2}{\sigma_0}  \overline{ n^2(x,t)}  (u-c)  $.  Multiplying it by the velocity $c$ and integrating over the system length one finds
\begin{align}
	\label{eq:W_dot}
	\dot{W} & = L \, \frac{e^2}{\sigma_0}\, 	\overline{ n^2} \, c \, (c-u). 
\end{align}
Dividing this by heat flux, $Ts_0 u$ one gets
\[
\mathrm{COP} = \mathrm{COP}_{\mathrm{C}} \, \frac{s_0 u \Delta T}{L \frac{e^2}{\sigma_0}	\overline{ n^2} c (c-u)},
\]
where  $\mathrm{COP}_{\mathrm{C}}=\frac{T}{\Delta T}$ is the Carnot efficiency. Expressing  $\Delta T$ in terms of $u$ using  Eq.~\eqref{eq:u_result} we get
\begin{align}
	\label{eq:COP_ratio}
	\frac{\mathrm{COP}}{\mathrm{COP}_{\mathrm{C}}} =  \frac{u}{c}     -   \alpha \, \frac{u^2}{c(c-u)},
\end{align}
where we introduced the dimensionless  disorder strength
\begin{align}
	\label{eq:alpha}
	\alpha = \frac{\langle (\delta n)^2 \rangle} {2\,  \overline{ n^2}}.
\end{align}
At a fixed temperature difference, the maximal COP in Eq.~\eqref{eq:COP_ratio}  is achieved at
\begin{align}
	\label{eq:u_optimal}
	\frac{u}{c} = 1 -\sqrt{\frac{\alpha}{1+ \alpha}}.
\end{align}
From Eq.~\eqref{eq:u_result}  it is easy to see that this occurs at
\begin{align}
	\label{eq:Delta_T_optimal}
	\Delta T = \frac{e^2 }{\sigma_0} \frac{c L  \overline{ n^2}}{s_0}\left[ \sqrt{\alpha (1+ \alpha) }  - \alpha \right].
\end{align}
The maximal value of COP is given by
\begin{align}
	\label{eq:COP_max}
	\frac{\mathrm{COP}_{\mathrm{m}}}{\mathrm{COP}_{\mathrm{C}}} &  =  \left(\sqrt{1+ \alpha}- \sqrt{\alpha}\right)^2 .
\end{align}
For weak disorder,  $\alpha \ll 1$, it nearly reaches the Carnot limit.  In this case  $u /c \approx 1$ in Eq.~\eqref{eq:u_optimal}, and  heat pumping is  almost perfect.

Equation \eqref{eq:W_dot} and subsequent results for the heat pump efficiency  may be also obtained by considering energy dissipation in the pump.  The considerations above Eq.~\eqref{eq:EMF_pumping} show that
within second order accuracy in $U/T$ the contributions of temperature gradients and viscous stresses to energy dissipation may be neglected. Thus the rate of energy dissipation per unit area  is given by $T\dot{s}=    k u^2  + \sigma_0 \mathcal{E}^2  $. Then using Eq.~\eqref{eq:EMF_pumping} we obtain
\begin{align}
	\label{eq:dissipation_u}
	T\dot{s}=  \frac{e^2}{\sigma_0} \left[  \frac{\langle (\delta n)^2 \rangle }{2}  u^2 +     n^2 (u-c)^2 \right].
\end{align}
 Note that the energy dissipation rate in the system is characterized by the intrinsic conductivity. This  reflects the fact that at small deviations from charge neutrality dissipation is caused by the electric fields arising in the electron liquid. The power consumed by the pump is given by sum of dissipated energy, $T L \dot{s}$ and heat flux $TJ_s$. This reproduces Eq.~\eqref{eq:W_dot}.

To summarize, a mechanism of electronic pumping of heat at charge neutrality was considered. Since the system is on average charge-neutral, the heat transfer proceeds at zero net charge current, and is not accompanied by voltage buildup along the system. This may prove advantageous for potential cooling applications (e.g. of microelectronic devices) where voltage buildup or presence of mechanically moving parts is undesirable.  Consideration focused on slow pumping in the regime of electron hydrodynamics. The key parameter of the pumping cycle is the dimensionless ratio of disorder to pumping strength ($\alpha$ in Eq.~\eqref{eq:alpha} in a traveling wave setup). At $\alpha \ll 1$ the heat flux is determined by the geometry of the pumping potential. For potentials in the form of a traveling wave the optimal efficiency of the pump,  Eq.~\eqref{eq:COP_max}, may come close to the Carnot limit. Equation~\eqref{eq:Delta_T_optimal} shows that optimal efficiency may be reached for a wide range of temperature differences by adjusting the pumping parameters.

I am grateful to V. Antonov, A. Barnard, D. Cobden, M. Feigel'man, P. Goldbart,  K. Matveev, M. Rudner, and B. Spivak for useful discussions.

\bibliography{heat_pump}

\end{document}